\documentclass[showpacs,amsmath,amssymb,prl,floatfix,twocolumn]{revtex4}
\usepackage{graphicx}
\usepackage{hyperref}
\usepackage{dcolumn}
\usepackage{bm}
\usepackage{epsfig}

\newcommand{\p}{\phi}

\begin{document}

\title{Statistical Approach to the Cosmological-Constant Problem on
    Brane Worlds}

\author{F.~K.~Diakonos and E.~N.~Saridakis
\footnote{E-mail: msaridak@phys.uoa.gr}} \affiliation{Department
of Physics, University of Athens, GR-15771 Athens, Greece}


\begin{abstract}
We investigate the physically accepted solutions of general
Braneworld scenarios, scanning uniformly the associated parameter
space. Without making any further assumptions we find that
solutions which give ``small'' Hubble parameters on the physical
brane, and therefore ``small'' effective cosmological constants on
the $4D$ Universe, are far more probable than those with ``large''
ones. Eventually, their distribution tends to the
$\delta$-function in the limit of continuous covering of the
parameter space.
\end{abstract}

\pacs{98.80.-k, 04.50.-h, 02.60.Cb}
 \maketitle

Why is the cosmological constant of the observable Universe many
orders of magnitude smaller than expected within the context of
ordinary gravity and quantum field theory? Since Einstein (1917)
many authors have tried to give a plausible and coherent answer
(\cite{Weinberg89} and references therein). Most of the earlier
attempts were within the general relativity context, while other
authors were led to the obscure anthropic principle
\cite{anthropic}. An outlet of different origin arose since Zumino
showed that unbroken supersymmetry leads to vanishing vacuum
energy \cite{Zumino75}. Unfortunately, and despite many attempts
in the context of supergravity (or non-conventional gravity)
\cite{DeWolfe99,sugra} and superstrings \cite{strings}, in the
real world we know of no mechanism that can produce an effective
cosmological constant within the remarkable small experimental
limits, and especially to keep it unaffected by quantum
fluctuations. Another interesting approach was followed by Hawking
and later by Coleman \cite{Hawking84}, showing that a distribution
of values for the effective cosmological constant with an enormous
peak at zero could arise. Although the whole approach raises many
doubts \cite{Weinberg89}, the idea of restricting the cosmological
constant to small values through statistical considerations is
interesting and consists the basis of the present work. Finally,
the approaches to the problem which remain in the conventional 4D
framework suffer from loss of generality \cite{conventional}, and
the use of cosmological observations on behalf of small values
pays the price of adding various non-general assumptions
\cite{Horava00}.

The discussion  was reheated after the arrival of observational
data  indicating that there is a positive cosmological constant in
the universe \cite{observ}. At the same time the Braneworld
scenario appeared in the literature \cite{RS99}. Amongst other
consequences, it offered new paths of producing a zero or small
(effective) cosmological constant on the physical brane consisting
our Universe \cite{braneworld,Forste00}, but none could offer a
robust and undoubted justification. Most of these works
re-introduced in the new Braneworld context (of various versions
and dimensions) the old fine-tuning procedure of conventional
cosmology.

In conclusion, there is not a convincing and general solution to
the cosmological constant problem. Definitely, a complete
apprehension will arise from a fundamental theory of nature and
the breaking of its symmetries. Since this theory is unknown at
present, effective or even phenomenological descriptions are
probably preferable, although incomplete. The minimum requirement
is that of remaining general and fine-tuning free. In this Letter
we follow this direction in the context of Braneworld scenario,
and we explore the physically accepted solutions in the parameter
space, without making any additional assumptions. Obviously this
approach cannot act as a solution to the cosmological constant
problem. However, it can be enlightening and provide some
arguments for its possible elaboration in statistical terms.

We consider a general class of Braneworld models, characterized by
the action:
\begin{eqnarray}
 &\kappa^2_5 S=\frac{1}{2}\int d^4xdy\sqrt{-g}\,R+\int
d^4xdy\sqrt{-g}\left[-\frac{1}{2}(\partial\p)^2-V(\p)\right]\nonumber\\
&-\sum_{i=1,2}\int_{b_i}d^4x\sqrt{-\gamma}\,\{[K]+U_i(\p)\},\ \ \
\ \  \label{action}
\end{eqnarray}
where $\kappa^2_5=\frac{1}{M^3_5}$ is a $5D$ gravitational
constant, and all quantities are measured in units of $M_5$. The
first term describes gravity in the five dimensional bulk space,
the second term corresponds to a minimally coupled bulk scalar
field with the potential $V(\p)$ and the last term corresponds to
two $3+1$ dimensional branes, which constitutes the boundary of
the $5D$ space. We include a potential term $U(\p)$ for the scalar
field at each of the two branes, denoting with $\gamma$ the
associated induced metric and with $K$ its extrinsic curvature.
The square brackets denote the jump of any quantity across a brane
($[Q]\equiv Q(y_{+})-Q(y_{-})$). Assuming $S^1/{\mathbb Z}_2$
symmetry across each brane we restrict our interest only in the
interbrane space \cite{Forste00,Tetradis01,brcod}, (we use
$[Q']_0=2Q'(0^{+}),\ [Q']_1=-2Q'(1^{-})$). As usual the two branes
are taken parallel, $y$ denotes the coordinate transverse to them
and isometry along three dimensional $\mathbf{x}$ slices including
the branes, is assumed. For the metric we choose the conformal
gauge \cite{brcod}:
\begin{equation}
ds^2=e^{2B(t,y)}\left(-dt^2+dy^2\right)+e^{2A(t,y)}d\mathbf{x}^2.
\label{metric}
\end{equation}
This metric choice, along with the residual gauge freedom
$(t,y)\rightarrow(t',y')$ which preserves its form, allows us to
``fix" the positions of the branes. Without loss of generality we
can locate them at $y=0,1$, having in mind that their physical
distance at a specific time is given by
$D(t)=\int_0^1dy\,e^{B(t,y)}$ \cite{brcod}. The non-trivial
five-dimensional Einstein equations consist of three dynamical:
\begin{eqnarray}\
\ddot{A}-A''+3\dot{A}^2-3A'^2=\frac{2}{3}e^{2B}V\nonumber\\
\ddot{B}-B''-3\dot{A}^2+3A'^2=-\frac{\dot{\p}^2}{2}+\frac{\p'^2}{2}-\frac{1}{3}e^{2B}V\nonumber\\
\ddot{\p}-\p''+3\dot{A}\dot{\p}-3A'\p'+e^{2B}V_{,\p}=0,
\label{eomfull}
\end{eqnarray}
and two constraint equations:
\begin{eqnarray}\
-A'\dot{A}+B'\dot{A}+A'\dot{B}-\dot{A}'=\frac{1}{3}\dot{\p}\p'\nonumber\\
2A'^2-A'B'+A''-\dot{A}^2-\dot{A}\dot{B}=-\frac{\dot{\p}^2}{6}-\frac{\p'^2}{6}-\frac{1}{3}e^{2B}V,\,
\label{constraints}
\end{eqnarray}
where primes and dots denote derivatives with respect to $y$ and
$t$ respectively. Additionally, from the boundary terms in the
action for the branes we obtain the following junction (Israel)
conditions:
\begin{eqnarray}\
[A']=\mp\frac{1}{3}\,Ue^B,\ \ [B']=\mp\frac{1}{3}\,Ue^B,\ \
[\p']=\pm e^B\,U_{,\p}, \label{junctions}
\end{eqnarray}
where the upper and lower signs refer to the branes at $y=0,1$,
respectively. For the bulk and brane potentials we assume:
\begin{equation}
V(\p)=\frac{1}{2}m^2\p^2+\Lambda,\ \
U_i(\p)=\frac{1}{2}M_i(\p_i-\sigma_i)^2+\lambda_i.
\label{branepot}
\end{equation}
This is a general form for $V(\p)$, consistent with the brane
stabilization mechanism \cite{Goldberger99}, $\Lambda$ being the
$5D$ bulk cosmological constant. For $U_i(\p)$ the quadratic form
is also general \cite{brcod}, incorporating the brane tensions
$\lambda_i$ as well as the value $\p_i$ of $\p$ on the i-th brane.
Finally, the induced $4D$ metrics of the two (``fixed''-position)
branes in the conformal gauge are simply given by:
$ds^2=-d\tau^2+a^2(\tau)\,d\mathbf{x}^2$, with $\tau_i=e^{B_i}t$
and $a_i=e^{A_i}$ the proper times and scale factors of the two
branes \cite{Saridakis:2007ns}. Thus, for the Hubble parameter on
the branes we acquire
\begin{equation}
H_i\equiv\frac{1}{a}\frac{da}{d\tau}\Big\vert_i
=e^{-B_i}\dot{A}_i, \label{Hubble}
\end{equation}
and we identify the physical brane with the one at $y=0$.

The described model is quite general and includes the full
spacetime evolution of the $5D$ Braneworld. In order to
investigate some of its consequences we first examine the
stationary sub-class of solutions, assuming that:
\begin{equation}
B(t,y)\rightarrow B(y),\ \ A(t,y)\rightarrow B(y)+H\,t.
\label{stationary}
\end{equation}
This choice leads to a fixed bulk geometry ($\p$ is time
independent), and maximally symmetric (de Sitter or Minkowski)
branes: $
ds^2=e^{2B(y)}\left(dy^2-dt^2+e^{2Ht}d\mathbf{x}^2\right)$, where
the Hubble parameter on the physical brane is simply
$H_0=e^{-B(0)}H$. If we set $V(\p)=0$ we can acquire analytical
solutions for the system of equations (\ref{eomfull}) and
(\ref{constraints}), depending on the sign of $H^2$. For $H^2>0$
and setting $H=|\sqrt{H^2}|$ we get:
\begin{eqnarray}
B(y)=B(0)+\frac{1}{3} \log{\left(\frac{B'(0)}{H}
\sinh{3Hy}+\cosh{3Hy}\right)} \nonumber\\
\p(y)=\p(0)-\frac{2}{\sqrt{3}}\log{\left[\frac{\left(e^{3Hy}-u\right)\left(1+u\right)}
{\left(e^{3Hy}+u\right)\left(1-u\right)}\right]}\  \label{pyp}
\end{eqnarray}
where $u=\sqrt{\frac{B'(0)-H}{B'(0)+H}}$, and always $|B'(0)|\geq
H$ for the existence of a solution. When $H^2<0$ and setting
$\theta=|\sqrt{-H^2}|$, the solutions can be obtained from
(\ref{pyp}) under the replacement $H\rightarrow i \theta$
\cite{Saridakis:2007cf}. We mention that we do not make additional
assumptions, excluding this case ($H^2<0$) by hand, since we
investigate any possible solution. Finally, expressions
(\ref{pyp}) have to satisfy the boundary conditions
(\ref{junctions}), and eventually $B(0)$, $B'(0)$ and $\p(0)$ and
$H$ are given in terms of the six parameters of the model
($M_0,\lambda_0,\sigma_0,M_1,\lambda_1,\sigma_1$).

In order for the solutions to be physically accepted, we have to
assure that no ``naked'' singularities are present in the
physically meaningful interbrane space, i.e between $y=0$ and
$y=1$. As it is shown in \cite{Tyewass01c}, the singularities of
the aforementioned solutions are indeed true (and not just
coordinate ones) since they possess incomplete geodesics reaching
a point of divergent curvature. An equivalent way to reveal the
true nature of these singularities, is by calculating explicitly
any of the 5D curvature invariants such are the Ricci scalar, the
Kretschmann scalar and the Weyl tensor, and showing that they
diverge. This has been performed in
\cite{DeWolfe99,brcod,Tyewass01c2}, and has been straightforwardly
verified by us that  these curvature invariants do diverge in the
singularities of the solutions (\ref{pyp}), offering a robust
proof of their true (naked) nature.

For $H^2<0$ case, the requirement of singularity absence forces
the physically accepted solutions to satisfy:
\begin{equation}
3\theta-\pi<\arctan\left(-\frac{\theta}{B'(0)}\right)<0.\label{wind2}
\end{equation}
Thus, this relation provides a narrow and absolute window for
$\theta$, excluding areas of the $6D$ parameter space leading to
$\theta$-values larger than $\pi/3$. For the $H^2>0$ case the
avoidance of singularities leads to:
\begin{equation}
H<|B'(0)|<H\coth(3 H),\label{wind1}
\end{equation}
and since $B'(0)$ and $H$ are functions of the six parameters of
the model, the window (\ref{wind1}) gives the limits of the
sub-space of the $6D$ parameter space that corresponds to accepted
solutions.
 Inequality (\ref{wind1}) becomes actually an equality for
large $H$ values ($H\gtrsim 2$ in the units we use) since both
edges practically coincide. The resulting relation acts as a
constraint in the $6D$ parameter space, and therefore the
parameter sub-space which corresponds to physically accepted
solutions with large  $H$ is of lower dimensionality, i.e it has
been reduced to a sub-surface:
\begin{equation}
H=|B'(0)|=H\coth(3 H).\label{wind1eq}
\end{equation}
In conclusion, from the above analytical calculation it is implied
that the demand for absence of divergences in the solutions
restricts the probable $H^2$ values (and therefore $H_0^2$ ones)
to a narrow window around zero (solutions with large $H$ have zero
measure). The additional requirement of fulfilling the junction
conditions on the second brane makes the parameter sub-space,
leading to a solution, in-homogenous and non-compact
\cite{Tetradis01}.

\begin{figure}[ht]
\begin{center}
\mbox{\epsfig{figure=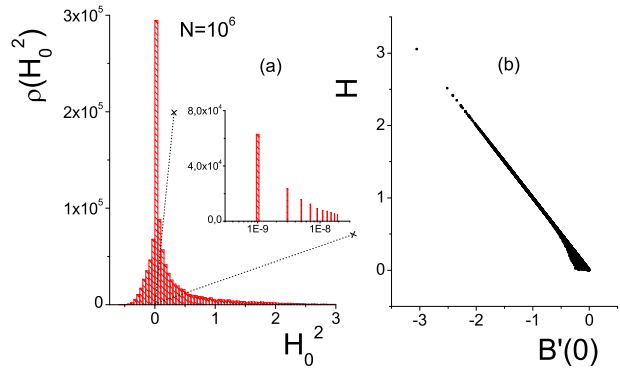,width=9cm,angle=0}} \caption{(Color
online) \it In graph a) we present the $H_0^2$ histogram for the
accepted solutions. In graph b) we depict $H$ vs $B'(0)$ for
solutions with $H^2>0$. Both graphs correspond to the stationary
case with zero bulk potential, and each edge of the parameter
space hyper-cube extends from -$10^3$ to $10^3$.} \label{zerobulk}
 \end{center}
 \end{figure}

Let us verify numerically the acquired results of the stationary
case with zero bulk potential. The main difficulty of solving the
equation system (\ref{eomfull})-(\ref{branepot}) is that one has
to satisfy the constraints (\ref{constraints}) and boundary
conditions (\ref{junctions}) at $t=0$, ensuring that no
divergencies are present in the interbrane space, task which
proves to be extremely hard in general. The restriction to $t=0$
is easily justified since the dynamical equations (\ref{eomfull})
preserve the constraints (\ref{constraints}), and furthermore for
the stationary case at hand the boundary conditions
(\ref{junctions}) are time-independent. The method we use is the
following: We first choose randomly the values of the 6 model
parameters, uniformly distributed in a 6D hyper-cube. The obtained
results do not depend on the hyper-cube's size, but on its
effectual covering (number of parameter multiplets used in the
calculation). We use the globally convergent Schmelcher-Diakonos
algorithm \cite{Diakonos} to solve the transcendental equation
system (\ref{constraints})-(\ref{junctions}), with accuracy
$10^{-13}$. Thus, all existing solutions are easily obtained,
which is not the case if one uses Newton's method \cite{brcod},
where a good initial guess, of a (relatively small) interval
containing the root to be found, is needed as an input. We find
hexads of parameters corresponding to acceptable solutions, and
calculate $H_0$ through $H_0=e^{-B(0)}H$. In fig.~\ref{zerobulk}a
we present the $H_0^2$ histogram for $N=10^6$ accepted solutions,
and we observe that $\rho(H_0^2)$ is strongly restricted around
zero according to windows (\ref{wind2}) and (\ref{wind1}).
 A numerical fit specifies that for $H_0^2>0$ the
aforementioned histogram is an exponential (coefficient of
determination $R^2\approx0.999$):
\begin{equation}
\rho(H^2_0)\sim e^{-\alpha H^2_0},\label{exponential}
\end{equation}
 with exponent $\alpha\approx3$.
 In order to provide an
additional verification, in fig.~\ref{zerobulk}b we depict $H$
versus $B'(0)$ for solutions that have $H^2>0$. We observe that
for small $H$ ($H<0.5$) inequality (\ref{wind1}) holds, and it
gradually transforms to equality (\ref{wind1eq}) as $H$ grows, i.e
the parameter sub-space that corresponds to accepted solutions
becomes thiner and thiner and finally results in a sub-surface.

We mention here that in our analysis we follow  the
``anti-fine-tuning'' approach to the problem, determining first,
unbiased (i.e randomly), the six model parameters and then
investigating the solution features. This is opposite to the usual
method in the literature, which is to fulfil the boundary
conditions in the first brane at $y=0$, choose an arbitrary  $H$,
solve the constraints (\ref{constraints}) towards $y=1$ and then
fit the brane parameters of the second brane so that the
corresponding junction conditions are fulfilled. Definitely this
fine-tuning procedure has the disadvantage of neglecting several
solutions, allowing for an apparent uniform distribution of the
$H$ value, contrary to the fact that the considered solution space
is a sub-surface of the full $6D$ parameter space. The source of
the misleading behavior is the over-determined nature of the
problem, i.e the existence of more relations and requirements to
be fulfilled (including conditions that are not expressible
through a strict equation, such is the absence of naked
singularities) than the number of model parameters. In such cases
our method is better in order to reveal the full characteristics
of the parameter space and its sub-space that corresponds to
solutions.

\begin{figure}[ht]
\begin{center}
\mbox{\epsfig{figure=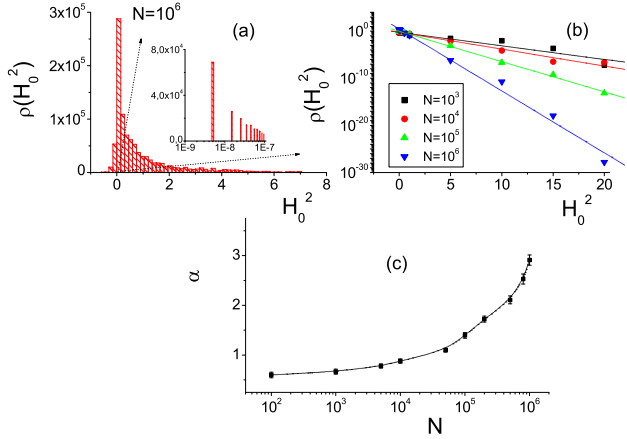,width=9cm,angle=0}} \caption{(Color
online) \it In graph a) we present the $H_0^2$ histogram for the
accepted solutions. In graph b) we show the histogram
$\rho(H_0^2)$ of $H_0^2$, for four different numbers of physically
accepted solutions, where the corresponding curves have been
normalized to area 1. The dotted lines correspond to linear fits.
In graph c) we depict the dependence of exponent $\alpha$, of the
exponential form (\ref{exponential}), on the physically accepted
solution number $N$. Numerical extrapolation gives
$\alpha\rightarrow\infty$ for $N\rightarrow\infty$. All graphs
refer to the stationary case with full bulk potential.}
\label{fullbulk}
\end{center}
\end{figure}

After this verification of the analytical calculations we can
proceed to numerical solutions for the stationary case with full
bulk potential. Again, we randomly choose  the values of the 8
model parameters ($M_0,\lambda_0,\sigma_0,M_1,\lambda_1,\sigma_1$,
$m$ and $\Lambda$), from a uniform distribution, and we solve the
transcendental equation system
(\ref{constraints})-(\ref{junctions}). We mention here that we do
not use any {\it {a priori}} restrictions on the parameter values
arising from physical arguments (for example $\Lambda<0$ in order
to acquire a $5D$ AdS bulk as required by many authors), since we
want to be as general as possible. In fig.~\ref{fullbulk}a we
present $\rho(H_0^2)$ for $N=10^6$ such solutions, and a numerical
fit for $H_0^2>0$ gives again an exponential ($R^2\approx0.998$)
of the form (\ref{exponential}) with exponent $\alpha\approx 2.9$.
Although an analytical treatment is not possible in this case, the
qualitative results are similar to those obtained for zero bulk
potential. Indeed, it seems that ``small'' $H_0$ arise from a
(non-homogenous and non-compact) sub-space of the parameter space,
which becomes thinner for ``larger'' $H_0$, resulting eventually
to a sub-surface.

The natural arising question is what factors affect the exponent
$\alpha$ of relation (\ref{exponential}). We can show numerically
that it is determined by the statistics and increases with the
number of solutions $N$. In fig.~\ref{fullbulk}b we depict
$\rho(H^2_0)$ vs $H^2_0$, for four different-$N$ cases. We clearly
see that the exponential form decays faster for larger $N$. In
fig.~\ref{fullbulk}c we present the values of the exponent
$\alpha$, arising from numerical fits of the form of
(\ref{exponential}), for various $N$ values. Indeed, $\alpha$
increases with $N$, and the increasing rate is robust although
quite slow. Numerical extrapolation gives
$\alpha\rightarrow\infty$ for $N\rightarrow\infty$. This is a
favoring result since it implies that covering continuously and
uniformly the parameter space, the physically accepted solutions
compose for $\rho(H_0^2)$ the $\delta$-function $\delta(H_0^2)$.
Solutions with large $H_0$ are not impossible, bur arise from a
sub-surface of smaller dimensionality. In the continuum limit,
large $H_0$ solutions are of zero measure while the ones with
$H_0\approx0$ are by far the most probable (note that although we
are using $M_5$ units, the presence of $\delta(H_0^2)$ leads to
$H_0\approx0$ in every units). As it is well known, $H_0^2$ in our
Universe is related to the (effective in a $5D$ model) $4D$
cosmological constant of our world. Since
$|\lambda_{\text{eff}}|\lesssim |H_0^2|$ we finally acquire
$\rho(|\lambda_{\text{eff}}|)\approx
\delta(|\lambda_{\text{eff}}|)$, compatible with observations.

\begin{figure}[ht]
\begin{center}
\mbox{\epsfig{figure=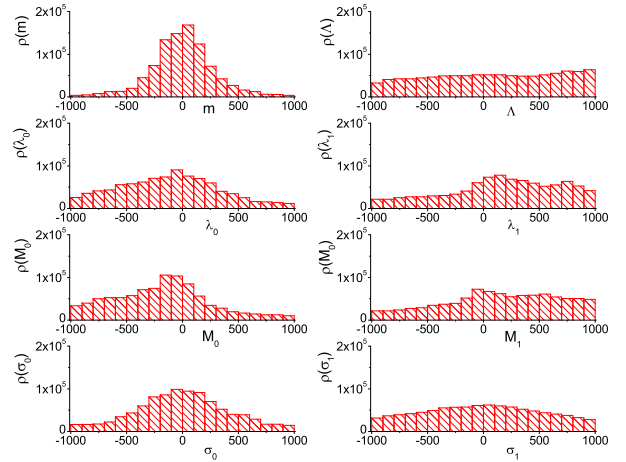,width=8cm,angle=0}} \caption{(Color
online) \it The histograms of the model parameters $m$, $\Lambda$,
$\lambda_0$, $\lambda_1$, $M_0$, $M_1$, $\sigma_0$, $\sigma_1$,
that correspond to $N=10^6$ accepted solutions, for the stationary
case with full bulk potential.} \label{paramhist}
\end{center}
\end{figure}

\begin{figure}[ht]
\begin{center}
\mbox{\epsfig{figure=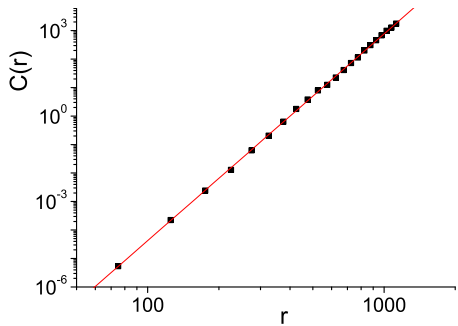,width=6cm,angle=0}} \caption{(Color
online) \it $C(r)$ versus $r$, in the region of intermediate
$r$'s. It follows the power-law form of (\ref{Crr}), with exponent
$\beta=7.2\pm0.3$.} \label{Cr}
\end{center}
\end{figure}

The distribution for $\lambda_{\text{eff}}$ obtained above is free
of both explicit and implicit fine-tuning, since we have not used
an explicit relation for $\lambda_{\text{eff}}$ in terms of the
model parameters and then force this expression to be small. On
the contrary we have just searched for general parameter octads
that correspond to solutions and for these octads we have
calculated $H_0$. In fig.~\ref{paramhist} we present the
histograms of the model parameters that correspond to $N=10^6$
accepted solutions, for the stationary case with full bulk
potential.
As we can clearly see, apart from a small tendency of $m$ to small
values, there is no specific preference to particular values. To
quantify this behavior we calculate the correlation dimension of
the sub-space of the parameter space that corresponds to accepted
solutions. Denoting by $\vec{r}_{ij}$ the 8D Euclidean distances
between two solution-octads $i$ and $j$:
\begin{eqnarray}
|\vec{r}_{ij}|^2=(m_i-m_j)^2+(\Lambda_i-\Lambda_j)^2+(\lambda_{0i}-\lambda_{0j})^2+\nonumber\\
+(\lambda_{1i}-\lambda_{1j})^2
+(M_{0i}-M_{0j})^2+(M_{1i}-M_{1j})^2+\nonumber\\+(\sigma_{0i}-\sigma_{0j})^2+(\sigma_{1i}-\sigma_{1j})^2,
\end{eqnarray}
we use the standard approach introduced by Grassberger and
Procaccia \cite{Grassberger} to calculate the correlation integral
\begin{equation}
C(r)=\frac{2}{N(N-1)} \sum_{i < j}^N \Theta(r-\vert \vec{r}_{ij}
\vert)\propto r^\beta\label{Crr},
\end{equation}
in order to obtain the correlation dimension $\beta$. In
fig.~\ref{Cr} we depict $C(r)$ versus $r$, in the region of
intermediate $r$ where the statistics allows for the power-law
form of relation (\ref{Crr}).
We observe that it clearly follows (\ref{Crr}) with exponent
$\beta=7.2\pm0.3$. Such a value (within statistical errors) is
very close to the value $\beta_{0}=8$ expected for a complete
covering of the 8D space. To estimate the effect of the finite
statistics we have repeated the calculations using the same number
of random octads within the corresponding hypercube (i.e without
taking into account any constraints induced by the requirement of
physically accepted solutions), and we have found a typical value
for the correlation dimension $\tilde{\beta}\approx 7.4\pm0.2$.
Therefore, our results should be compared with this value and not
with the value $\beta_{0}=8$. Thus, based on this fact, we safely
conclude that the obtained $H^2_0$-histogram and
$\lambda_{\text{eff}}$-distribution, are fine-tuning-free.

In the above analysis we have been restricted to stationary
solutions of the form (\ref{stationary}) without bothering about
time evolution, since in this case $H_0=H_0(0)=const.$ by
construction. The question is what can be said about the full
dynamics of equations (\ref{eomfull}) and (\ref{constraints}),
where $H_0$ can be varying. Fortunately, the solutions of the full
dynamics seem to consist of stationary or slowly varying ones
\cite{brcod,Tetradis01}. In such an evolution $H_0(t)$ can change
by at most one order of magnitude. This behavior justifies the
above stationary treatment, since the obtained results for $t=0$
will not change qualitatively for $t>0$.

In order to present this behavior in a more transparent way we
investigate numerically the full dynamical system (\ref{eomfull}),
(\ref{constraints}) and (\ref{junctions}), that is without
imposing the ansatz (\ref{stationary}). In particular, we first
perform the procedure described above for $t=0$ (or equivalently
for  initial brane proper-time $\tau=0$), in order to extract
$N=5\times 10^4$ parameter octads that correspond to solutions,
and for these octads we calculate $H_0(0)$ and
$\rho\left(H_0^2(0)\right)$. For these parameter octads we evolve
the system according to (\ref{eomfull}) and (\ref{junctions}), and
thus for each brane proper-time we obtain the corresponding
histogram $\rho\left(H_0^2(\tau)\right)$. We mention that for a
significant portion of the initially acquired solutions
($\approx20\%$), the subsequent time evolution leads to
divergences, since the singularities transit from outside to
inside the interbrane space. This behavior was also found in
\cite{brcod}. Thus these parameter octads are neglected.

\begin{figure}[ht]
\begin{center}
\mbox{\epsfig{figure=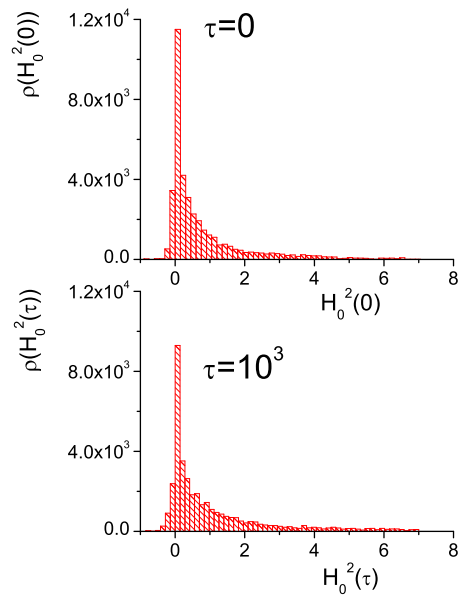,width=6cm,angle=0}} \caption{(Color
online) \it  The initial histogram $\rho\left(H_0^2(0)\right)$
(upper graph) and the histogram $\rho\left(H_0^2(\tau)\right)$ for
$\tau=10^3$ in $M_5$ units (lower graph), for $N=5\times 10^4$
physically accepted solutions of the full dynamical system
(\ref{eomfull}), (\ref{constraints}) and (\ref{junctions}), with
full brane and bulk potentials.} \label{tauhist}
\end{center}
\end{figure}

\begin{figure}[ht]
\begin{center}
\mbox{\epsfig{figure=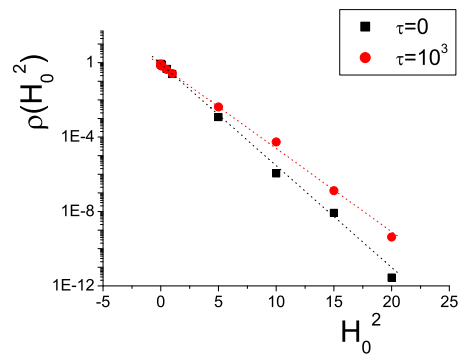,width=6cm,angle=0}} \caption{(Color
online) \it  The initial histogram $\rho\left(H_0^2(0)\right)$ and
the histogram $\rho\left(H_0^2(\tau)\right)$ for $\tau=10^3$ in
$M_5$ units, for $N=5\times 10^4$ physically accepted solutions of
the full dynamical system (\ref{eomfull}), (\ref{constraints}) and
(\ref{junctions}), with full brane and bulk potentials. The
histograms have been normalized to area 1 and the dotted lines
correspond to linear fits.} \label{tauslope}
\end{center}
\end{figure}

In fig.~\ref{tauhist} we present the initial histogram
$\rho\left(H_0^2(0)\right)$ (upper graph) and the histogram
$\rho\left(H_0^2(\tau)\right)$ for $\tau=10^3$ in $M_5$ units
(lower graph).
As we observe, indeed the variation of $H_0(\tau)$ from its
initial value $H_0(0)$ is less than an order of magnitude for the
significant portion of cases. Note also that this variation can be
either an increase or  a decrease. Thus, in summary, the final
histogram is only slightly wider. To quantify this deformation, in
fig.~\ref{tauslope} we present the aforementioned histograms in
logarithmic scale, normalized to area 1.
As we can see, time evolution does not change our results
significantly. Thus, in the limit $N\rightarrow\infty$ the
$\delta$-function-form $\rho(|\lambda_{\text{eff}}|)\approx
\delta(|\lambda_{\text{eff}}|)$ will be preserved.

One enlightening interpretation of the aforementioned result is
obtained by a parallelism with the equilibration mechanism of an
isolated macroscopic system through entropy maximization.
"Peculiar" states are not impossible, but there is extremely small
probability for their realization. In a similar way, in braneworld
models, the large majority (in the continuum limit ``almost'' all)
of the parametric combinations that correspond to physically
accepted solutions, lead to an extremely small cosmological
constant.

In conclusion, using a statistical approach, we have shown that in
the context of a general Braneworld scenario the acceptable
solutions of the Einstein equations are compatible with a
$\delta$-function distribution for the $4D$ effective cosmological
constant. Additionally, within this treatment,
$\lambda_{\text{eff}}$ is small and remains so even if the model
parameters vary by many orders of magnitudes due to successive
cosmological phase transitions. Of course the above analysis
cannot at all act as a solution to the cosmological constant
problem. It only provides some argument for its possible
elaboration in statistical terms.

\vskip .1in \noindent {\bf {\large Acknowledgment}}

We thank G. Felder, A. Frolov, G. Kofinas and N. Tetradis for
useful discussions.


\begin{thebibliography}{9}

\bibitem{Weinberg89}
 S.~Weinberg, Rev. Mod. Phys. {\bf 61}, 1 (1989).

\bibitem{anthropic}
 A.~Vilenkin, Contribution to
{\it  Universe or Multiverse}, Cambridge University Press (2007).


\bibitem{Zumino75}
B.~Zumino, Nucl. Phys. B {\bf 89}, 535 (1975).


\bibitem{sugra}
Z.~Kakushadze, Nucl. Phys. B {\bf 589}, 75 (2000);
Y.~Aghababaie,
C.~P.~Burgess, S.~L.~Parameswaran and F.~Quevedo, Nucl. Phys. B
{\bf 680}, 389 (2004);
A.~J.~Tolley, C.~P.~Burgess, D.~Hoover and
Y.~Aghababaie, JHEP {\bf 0603}, 091 (2006).


\bibitem{DeWolfe99}
O.~DeWolfe, D.~Z.~Freedman, S.~S.~Gubser and A.~Karch,  Phys. Rev.
D {\bf 62}, 046008 (2000).

\bibitem{strings}
E.~Witten,  {\it{Sources and detection of dark matter and dark
energy in the universe}}, Marina del Rey 2000;
 R.~Bousso and J.~Polchinski,
JHEP {\bf 0006}, 006 (2000).


\bibitem{Hawking84}
S.~W.~Hawking, Phys. Lett. B {\bf 134}, 403 (1984);
 S.~Coleman,
Nucl. Phys. B {\bf 310}, 643 (1988).



\bibitem{conventional}
I.~P.~Neupane, B.~M.~N.~Carter, Phys. Lett. B {\bf 638}, 94
(2006).

\bibitem{Horava00}
 H.~Martel,
P.~R.~Shapiro and S.~Weinberg, Astrophys. J. {\bf 492}, 29 (1998).

\bibitem{observ}
 S.~Perlmutter {\it et al.}, Astrophys. J. {\bf
517}, 565 (1999).


\bibitem{RS99}
L.~Randall and R.~Sundrum, Phys. Rev. Lett. {\bf 83}, 3370 (1999);
 L.~Randall and R.~Sundrum, Phys. Rev.
Lett. {\bf 83}, 4690 (1999).


\bibitem{braneworld}
 S.~Kachru, M.~Schulz and E.~Silverstein, Phys. Rev. D {f\bf 62}, 045021
(2000);
 S.~P.~de Alwis, Nucl. Phys. B {\bf 597},
263 (2001);
 S.~H.~Henry Tye and
I.~Wasserman, Phys. Rev. Lett. {\bf 86}, 1682 (2001);
 A.~Kehagias and K.~Tamvakis, Mod. Phys.
Lett. A {\bf 17}, 1767 (2002);
 S.~Mukohyama and L.~Randall, Phys. Rev. Lett. {\bf 92}, 211302
(2004).

\bibitem{Forste00}
S.~Forste {\it et al.}, Phys. Lett. B {\bf 481} 360 (2000);
S.~Forste {\it et al.}, JHEP {\bf 0009}, 034 (2000).

\bibitem{Tetradis01}
N.~Tetradis, Phys. Lett. B {\bf 509}, 307 (2001).


\bibitem{brcod}
J.~Martin, G.~N.~Felder, A.~V.~Frolov, M.~Peloso and L.~A.~Kofman,
Phys. Rev. D {\bf 69}, 084017 (2004).


\bibitem{Goldberger99}
W.~D.~Goldberger and M.~B.~Wise, Phys. Rev. Lett. {\bf 83}, 4922
(1999).

\bibitem{Saridakis:2007ns}
  E.~N.~Saridakis,
  JCAP {\bf 0804}, 020 (2008).


\bibitem{Saridakis:2007cf}
  E.~N.~Saridakis,
  Nucl.\ Phys.\  B {\bf 808}, 224 (2009).


\bibitem{Tyewass01c}
 P.~Kanti, K.~A.~Olive
and M.~Pospelov, Phys. Lett. B {\bf 481}, 386 (2000).

\bibitem{Tyewass01c2}
E.~Flanagan, S.-H.~H.~Tye and I.~Wasserman, Phys. Lett. B {\bf
522}, 155 (2001).

\bibitem{Diakonos}
P.~Schmelcher and F.~K.~Diakonos, Phys. Rev. Lett. {\bf 78}, 4733
(1997); F. K. Diakonos, P. Schmelcher and O. Biham, Phys. Rev.
Lett. {\bf 81}, 4349 (1998).

\bibitem{Grassberger}
P.~Grassberger and I.~Procaccia, Physica D {\bf 9}, 189 (1983).



\end{thebibliography}
\end{document}